\documentclass[prb,preprint]{revtex4-1} 

\usepackage{amsmath}
\usepackage{graphicx}
\usepackage[section]{placeins}
\usepackage{soul}

\begin{document}

\title{Discretely structured magnetic flux concentrators}

\author{Yujun Shi}
\email{yujunshi@sxu.edu.cn}
\affiliation{Collage of Physics and Electronics Engineering, Shanxi University, Taiyuan 030006, China}

\author{Xiaoting Feng}
\affiliation{Collage of Physics and Electronics Engineering, Shanxi University, Taiyuan 030006, China}

\begin{abstract}
Conventional magnetic flux concentrators (MFCs) are typically designed with solid structures, which may not be optimal for applications requiring lightweight design or material efficiency. In this study, we investigate the feasibility of spatially discretized MFCs in guiding and concentrating magnetic flux, through finite element simulations. Our results demonstrate that discretely structured MFCs can achieve performance comparable to that of solid counterparts while significantly reducing material usage.  For instance, even with a non-optimized discretized design, discretization along a single dimension can reduce material usage by an order of magnitude. Moreover, for three-dimensional structural devices, discretization along two dimensions has the potential to reduce material consumption by two orders of magnitude. This may offer a notable advantage in applications where weight reduction and cost efficiency are of primary concern.
\end{abstract}

\maketitle

\section{Introduction}
A magnetic flux concentrator (MFC) is a device or material structure designed to direct, concentrate, or amplify magnetic flux from one region to another. It essentially focuses magnetic field lines in a specific area, which is particularly useful in applications requiring an intensified magnetic field without increasing power input, such as weak magnetic field sensing \cite{1634424,10.1063/1.3056152,PhysRevResearch.2.023394}. With the exception of certain designs utilizing the diamagnetism of superconductors \cite{5109580,ZHANG20111547}. MFCs are typically made from high magnetic permeability materials, such as soft ferromagnetic alloys or composite materials. In traditional designs, MFCs often employ continuous solid structures, such as cones or rods, to guide and concentrate magnetic flux. However, when magnetic field lines in a region of free space are already attracted by nearby high-permeability materials, leading to a sparser distribution of field lines, adding more high-permeability materials to this space becomes unnecessary from the perspective of efficiently guiding magnetic field lines. Based on this simple consideration, one can achieve a similar concentration effect using spatially dispersed structures, such as arrays of discrete high-permeability elements or engineered composites. These structures guide the magnetic field through the air gaps or voids between the elements, rather than relying on a solid mass of material.

In fact, the use of discrete structures to implement magnetic flux concentrators (MFCs) has been reported, particularly in the metamaterial designs developed over the past two decades based on transformation optics. For example, Carles Navau et al. (2012) introduced a cylindrical metamaterial shell using a nonlinear space transformation \cite{PhysRevLett.109.263903}. This shell exhibited extreme radial and angular permeability components, $\mu_\rho\to\infty$ and $\mu_\theta\to0$, enabling it to either concentrate an externally applied static magnetic field within its interior or expel the field of an interior source to the exterior. The cylindrical shell was realized using alternating, discretely distributed pieces of superconductor (SC) and ferromagnetic (FM) materials, which provided a good approximation of the required anisotropic permeability components. Their subsequent experiments \cite{10.1063/1.4903867} demonstrated that even a cylindrical shell composed solely of alternating, discretely distributed soft ferromagnetic (FM) pieces could achieve similar MFC functionality.

However, in traditional electromagnetism and electrical engineering, detailed discussions on discretely structured magnetic flux concentrators (MFCs) remain relatively rare. This paper aims to investigate, through finite element simulations, the effectiveness of discretely structured MFCs in guiding or concentrating magnetic flux relative to their solid counterparts. Simulation results indicate that discretely structured MFCs can achieve performance comparable to that of solid structures while significantly reducing material usage. This offers a notable advantage in applications prioritizing lightweight design or cost efficiency.

All numerical simulations in this paper were conducted using the MagnetoDynamics2D module of Elmer, which is an open-source finite element software for multiphysical problems \cite{elmer}. To simplify the simulation, we model MFCs in two dimensions, which corresponds to assuming an infinitely long third dimension in the three-dimensional structure. The organization of the paper is as follows: In Section 2, we define the discrete unit as a rectangular strip and simulate its effect on flux lines and field density when placed in a uniform external magnetic field, which is parallel to the long side of the strip. We investigate the magnetic flux concentration as a function of the width-to-length ratio. The conclusion drawn is that the smaller the width-to-length ratio (i.e., the narrower the rectangular strip), the higher the efficiency of magnetic flux convergence around the strip. This finding forms the basis for the discretization of MFCs. In Section 3, we take a strip with a width-to-length ratio \( d/L = 1/60 \) as the discrete element, model the solid rectangular structure discretely, and simulate the magnetic field concentration effect when the discrete structure and its corresponding solid structure are placed in a uniform external field. In Section 4, we discretize a typical MFC structure and simulate its magnetic field concentration effect when placed in an external uniform magnetic field, as well as the expulsion of the internal magnetic field to the exterior when magnetic sources are placed inside. The simulation results in both Sections 3 and 4 indicate that discretely structured MFCs can achieve performance comparable to that of solid structures while significantly reducing material usage. Finally, we discuss potential applications for discrete MFC structures in the conclusion section.

\section{A rectangular strip in a uniform external magnetic field}
We consider the discrete unit to be a rectangular strip composed of a homogeneous material with an isotropic relative permeability of $\mu = 100$ (unless otherwise specified, all magnetic materials in this study are assumed to have a relative permeability of 100). The long edge of the strip is fixed in length and oriented parallel to the direction of the uniform external magnetic field $B_0$.

\begin{figure}[h!]
	\centering
	\includegraphics[width = 0.8\textwidth]{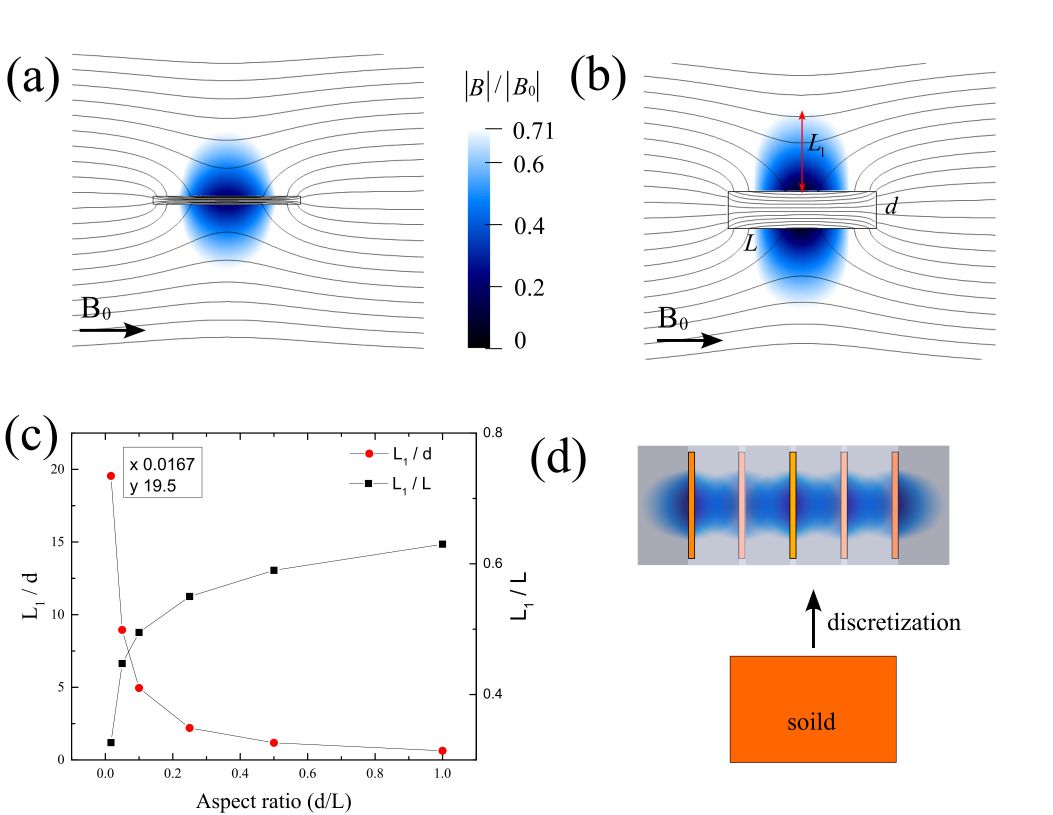}
	\caption{(a, b) Distribution of magnetic flux around a single rectangular strip with aspect ratios $d/L = 0.1$ and $d/L = 0.25$, respectively. The relative magnetic energy density in the blue gradient areas on both sides of the strip decreases to below $1/2$. (c) The dependence of $L_1/d$ and $L_1/L$ on the aspect ratio $d/L$, $L_1$ represents the maximum distance between the boundary of the gradient region and the surface of the rectangular strip. (d) Illustration of a discretization rule where unit rectangular strips are placed at intervals of $L_1$ along the short edge.
	}
	\label{a single rectangular strip}
\end{figure}

Figure \ref{a single rectangular strip} presents the simulation results of the magnetic flux density distribution. Evidently, due to the high magnetic permeability of the material, the magnetic field lines near both sides of the rectangular strip are drawn inward, resulting in reduced magnetic flux density in these regions. In Figures \ref{a single rectangular strip}(a) and (b), the blue gradient areas on both sides of the strip correspond to a relative magnetic flux density magnitude ($|B|/|B_0|$) distribution ranging from 0 to $\sqrt{2}/2$. In these magnetic flux depleted regions, the relative magnetic energy density drops below $1/2$. As shown in Figure \ref{a single rectangular strip}(b), we define $L_1$ as the maximum distance between the boundary (where $|B|/|B_0|=\sqrt{2}/2$) of the gradient region  and the surface of the rectangular strip. $L_1$ can be used to characterize the concentration effect of rectangular strip on magnetic flux, or more accurately, the effect of depleting the surrounding magnetic flux. Figure \ref{a single rectangular strip}(c) illustrates the dependence of $L_1/d$ and $L_1/L$ on the aspect ratio $d/L$, where $L$ and $d$ are the length and width of the rectangular strip, respectively. Since $L$ is fixed, $L_1/L$ represents the absolute length of $L_1$, while $L_1/d$ represents its relative length with respect to the strip width. We observe that $L_1/L$ decreases as the aspect ratio decreases, indicating that a narrower rectangular strip results in a smaller absolute width of the region where the relative magnetic energy density falls below $1/2$. Conversely, $L_1/d$ increases significantly as the aspect ratio decreases, suggesting that a narrower rectangular strip leads to a larger relative width (normalized by $d$) of the region where the relative magnetic energy density falls below $1/2$.

Assuming a discretization rule where unit rectangular strips are placed at intervals of $L_1$ along the short edge, the significant increase in $L_1/d$ as the aspect ratio decreases implies that the material proportion in the spatial volume decreases as the unit strip becomes thinner. This suggests that when discretizing a continuous three-dimensional structure using thin sheets in one direction, thinner unit sheets lead to more effective discretization.

\section{Discretization of rectangular structures}
Next, we investigate the specific effects of employing discrete structures in MFCs compared to solid structures in guiding or concentrating flux via simulations. First, we examine the simplest case: a solid rectangle and its discretized equivalent.

\begin{figure}[h!]
	\centering
	\includegraphics[width = 0.8\textwidth]{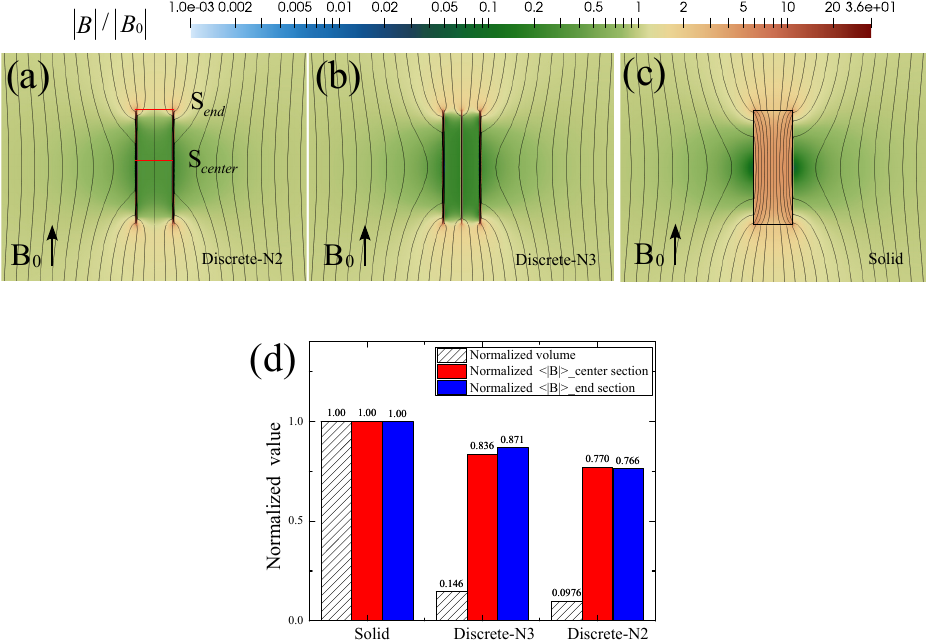}
	\caption{(a, b, c) Distribution of magnetic flux around two discrete rectangular structures and their corresponding solid rectangular structures. (d) The normalized values of the mean magnetic flux density magnitudes on the center and end cross-sections of the three structures, along with their corresponding material volumes.}
	\label{Magnetic field of rectangle}
\end{figure}

As shown in Figure \ref{Magnetic field of rectangle}(a), the discretely structured MFC consists of two rectangular strips, both with an aspect ratio of 1/60, denoted as Discrete\_N2. According to the discretization rules in Section~2, the center-to-center distance between the two strips set to $L_1$, which equals to 19.5 times their width. The corresponding solid rectangular structure is shown in Figure \ref{Magnetic field of rectangle}(c), with an aspect ratio of  20.5/60 (approximately 1/3). Figure \ref{Magnetic field of rectangle}(b) illustrates an extended version of Discrete-N2, where an additional strip is introduced, denoted as Discrete-N3. The influence of these two discrete structures on the magnetic flux density distribution in the surrounding space closely resembles that of the solid structure. Figure \ref{Magnetic field of rectangle}(d) presents the normalized mean magnetic flux density magnitudes at the center and end cross-sections of the three structures, along with the corresponding material volumes. The data for the solid structure are used for normalization. The material volume of Discrete-N2 is approximately 0.1, achieving normalized $<|B_{center\_section}|/|B_0|> \sim 0.77$ and $<|B_{end\_section}|/|B_0|>\sim 0.77$. Whereas Discrete-N3 requires a material volume of approximately 0.15 to achieve better magnetic flux concentration, with normalized values of 0.84 and 0.87 at the center and end sections, respectively. This indicates that discrete structures perform similarly to solid structures in concentrating magnetic flux. By employing thinner rectangular strips as discrete elements and optimizing the design, the discretized rectangular structure can achieve maximal efficiency in guiding or concentrating external magnetic flux.

\section{Discretization of typical MFC structures}

\begin{figure}[h!]
	\centering
	\includegraphics[width = 0.65\textwidth]{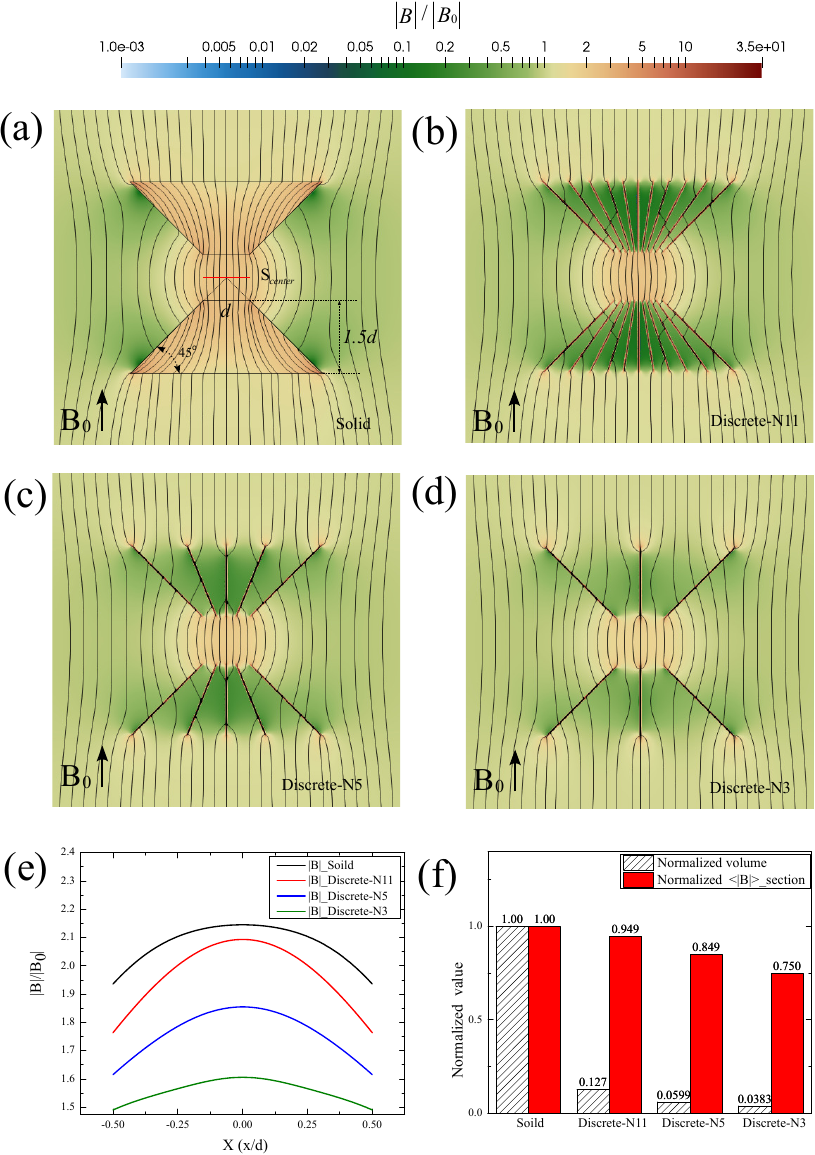}
	\caption{Magnetic flux concentration. (a, b, c, d) Distribution of magnetic flux around three discrete MFCs and a solid MFC. (e) The relative magnetic flux density distribution on the center cross-section of the four MFCs. (f) The normalized values of the mean magnetic flux density magnitudes on the center cross-sections of the four MFCs, along with their corresponding material volumes.
	}
	\label{Magnetic flux concentration}
\end{figure}

We begin to discretizie a typical MFC structure and simulate its effect on the surrounding magnetic flux distribution. The solid structure of the MFC consists of two identical isosceles trapezoids with their top edges facing each other, as shown in Figure \ref{Magnetic flux concentration}(a). The geometric parameters of the trapezoid are as follows: the top edge length is $d$, the height is $1.5~d$, the base angle is $45^\circ$, and the distance between the top edges of the two trapezoids is also $d$. Figures \ref{Magnetic flux concentration}(b–d) present three structures discretizied along angular direction, with varying spacings. We denote these structures as Discrete-N11, Discrete-N5, and Discrete-N3, respectively. The discrete units are rectangular strips with an aspect ratio of 1/60.

\subsection{ Concentration of an external field in the interior of MFCs}
Figure 4 presents the magnetic flux density concentration in the central region of these MFCs when placed in a uniform external magnetic field. It is evident that, similar to the solid structure, the three discretely structured MFCs are also able to guide and concentrate magnetic flux lines within their internal regions. Figure 4(f) shows the normalized values of the mean magnetic flux density magnitudes at the center cross-sections of the four MFCs. Similar to the rectangular structure discussed in Section~3, it can be observed that Discrete-N11 achieves a relative magnetic flux density of 0.95 at the center cross-section, while the required relative volume of magnetic material is only 0.13.

\subsection{Expulsion of an interior field to the exterior of MFCs}

Static magnetic fields diminish rapidly as the distance from their source increases. For a magnetic dipole, the magnetic field strength \( B \) falls off as the inverse cube of the distance. When a magnetic field source is placed inside an MFC, the MFC may enhances the magnetic field strength in the external far-field region. Alternatively, it can be said that the MFC expels part of the internal magnetic field outward, thereby enabling the extension of the static magnetic field over a finite distance.

\begin{figure}[h!]
	\centering
	\includegraphics[width = .8\textwidth]{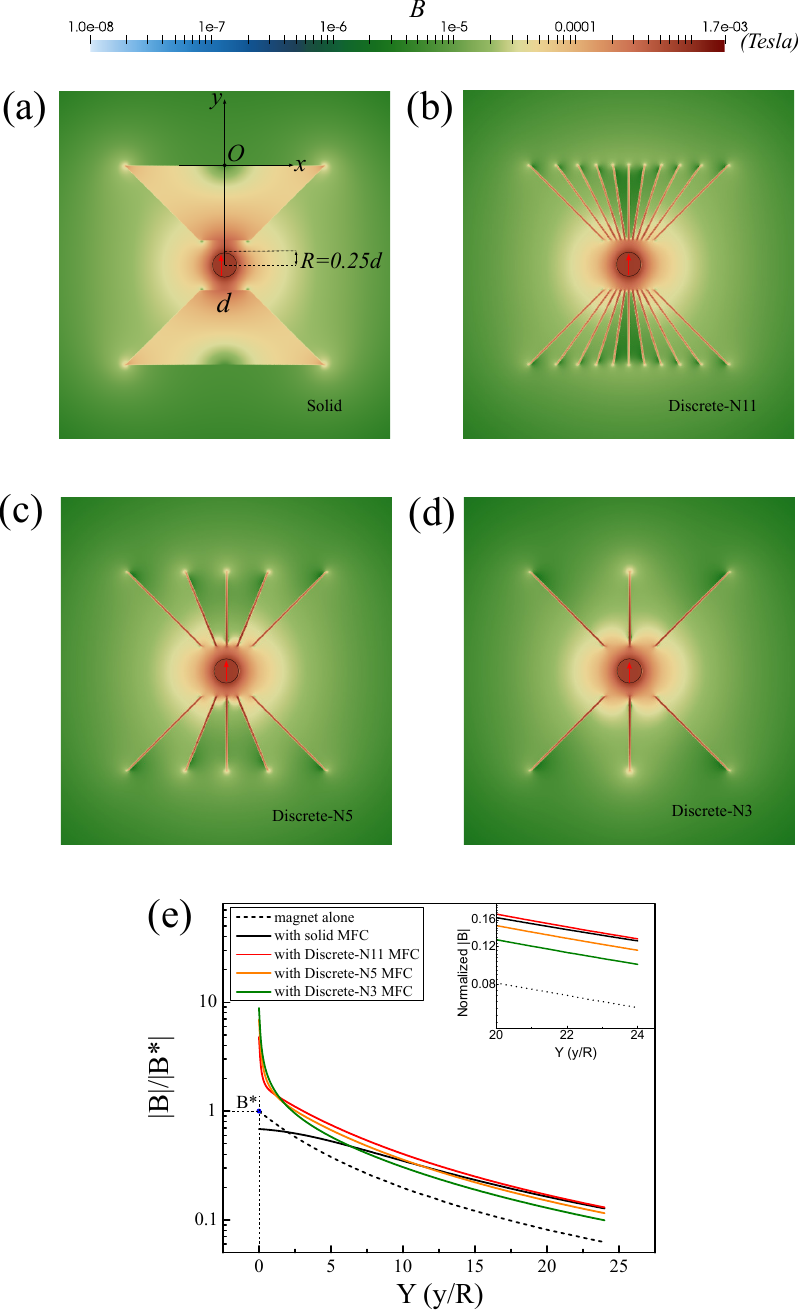}
	\caption{Magnetic flux expulsion. (a, b, c, d) Distribution of magnetic flux around three discrete MFCs and a solid MFC, when a circular magnet is placed at the center of their interior, with magnetization direction along the y-axis as shown in Figure (a). (e) The relative magnetic flux density distribution along the positive y-axis, with the coordinate system defined as shown in figure(a). }
	\label{Magnetic field expulsion}
\end{figure}

In our study, we placed a uniformly magnetized circular magnet (with a radius of \( R = 0.25~d \)) in the central region of the MFCs. The magnetization direction of the magnet is aligned with the central axis of the MFC. Figure \ref{Magnetic field expulsion} shows the corresponding magnetic field distribution. When considering only the magnetic field distribution in the external space of the MFC, as illustrated in Figure \ref{Magnetic field expulsion}(a), the magnetic field strength \( B \) varies along the central axis with distance \( y \) from the MFC surface, as shown in Figure \ref{Magnetic field expulsion}(e). The magnetic field data presented has been normalized by the field strength at \( y = 0 \), when the bare magnet is located without MFCs. Disregarding the significant variations in the magnetic field as $y$ approaches 0, i.e., near the surface of the MFC, the MFC structure enhances the magnetic field strength at greater distances. For instance, at \( y = 20R \), the magnetic field gain for the solid structure MFC is approximately 2. The discrete MFC shows a similar gain, especially for the denser Discrete-N11 structure, which even exhibits a slightly higher gain factor than the solid structure.

\section{Conclusion}

In summary, our simulations confirm that discretely structured MFCs can achieve performance comparable to that of solid structures while significantly reducing material usage. In the two-dimensional simulations, we discretized the structure along only one dimension, reducing the proportion of the high-permeability material to as low as 1/10. For a truly three-dimensional structure, discretization can be applied in two dimensions, leading to the expectation that material usage could potentially be reduced by two orders of magnitude without significantly degrading MFC performance.

Furthermore, although our discussion of MFCs has been limited to static magnetic fields, previous studies have shown that MFCs exhibit similar flux concentration effects under low-frequency alternating magnetic fields (up to tenths of kilohertz) \cite{10.1063/1.4903867, Kibret_2016}. The specific efficacy of discretely structured MFCs under such low-frequency alternating fields remains an open question for future investigation. Additionally, we note that Bernhard Gleich et al. (2023) reported in \textit{Science} on the development of miniature magneto-mechanical resonators with volumes below 1 cubic millimeter for wireless tracking and sensing \cite{science.adf5451}. These resonators utilize the rotational oscillation of levitated micro-magnetic spheres at frequencies of several kilohertz. Through electromagnetic excitation and measurement, they demonstrated real-time tracking of the position and orientation of a flying bee within a spatial range of approximately 20 cm. This suggests potential applications in low-cost wireless tracking, particularly in medical contexts. Integrating a discretized MFC around such lightweight magneto-mechanical resonators may further expand their spatial range for wireless tracking and sensing applications.

\section*{Acknowledgments}
This work has been funded by Research Proiect Supported by Shanxi Scholarship Council of China (Grant No. 2023-033), and the Fundamental Research Program of Shanxi Province (Grant No. 202303021221071).

\section*{Conflict of Interest}
The authors have no conflicts of interest to disclose.

\FloatBarrier

\bibliographystyle{unsrt}
\bibliography{reference.bib}

\end{document}